%% file: sample-authordraft.tex
\documentclass[sigconf,authorversion,nonacm]{acmart}
\usepackage{mathtools}
\usepackage{bbm}
\usepackage{verbatim}
\usepackage{enumitem}
\usepackage[flushleft]{threeparttable}
\usepackage{subcaption}
\usepackage{xcolor,colortbl}
\usepackage{caption} 

\AtBeginDocument{%
  \providecommand\BibTeX{{%
    \normalfont B\kern-0.5em{\scshape i\kern-0.25em b}\kern-0.8em\TeX}}}

\setcopyright{none} 

\begin{document}

\title{Treatment Effect Detection with Controlled FDR under Dependence for Large-Scale Experiments}

\author{Yihan Bao, Shichao Han, Yong Wang}




\begin{abstract}
  Online controlled experiments (also known as A/B Testing) have been viewed as a golden standard for large data-driven companies since the last few decades. The most common A/B testing framework adopted by many companies use "average treatment effect" (ATE) as statistics. However, it remains a difficult problem for companies to improve the power of detecting ATE while controlling "false discovery rate" (FDR) at a predetermined level. One of the most popular FDR-control algorithms is Benjamini–Hochberg (BH) method, but BH method is only known to control FDR under restrictive positive dependence assumptions with a conservative bound. In this paper, we propose statistical methods that can systematically and accurately identify ATE, and demonstrate how they can work robustly with controlled low FDR but a higher power using both simulation and real-world experimentation data. Moreover, we discuss the scalability problem in detail and offer comparison of our paradigm to other more recent FDR control methods, e.g., knockoff, AdaPT procedure, etc.
\end{abstract}



\maketitle

\section{Introduction}
 Online controlled experiments, also known as A/B testing, are proven to embody the best scientific design for establishing the casual relationship between treatment effect and users' observable behavior \cite{kohavi2009controlled}. To better accommodating users and improving product ideas, many IT companies have built their own in-house online experimentation platforms to meet their complex experimentation needs, e.g., Facebook \cite{bakshy2014designing}, LinkedIn \cite{xu2015infrastructure}, and Twitter \cite{al2011experimental}.
 
 At our company, we have also seen significant growth in A/B testing usage over the past two years. There are more than a billion users participating in thousands of experiments simultaneously, and with statistical inference and estimation conducted to thousands of online metrics, ranging from app performance metrics, customer engagement metrics, to data quality metrics (such as missing log).
 
One of the important reasons for carrying out A/B tests is to measure how a certain feature (treatment) will impact the metrics we are interested in. In order to formalize the problem that can be better equipped with statistical tools, we can define each metric under Rubin's causal framework \cite{rubin1974estimating}, and the difference before and after receiving treatment on each metric across individuals can be defined as "average treatment effect" (ATE).

However, with thousands of user characteristics available to IT companies, in each A/B testing, practitioners tend to test at least tens of metrics concurrently, hoping to discover how the new features and strategies they propose will significantly change users' behavior in different aspects. This approach will lead to a strong threat from false discoveries, which is a statistical artifact known as "multiple comparison". If we follow this "naive" approach by simply computing the estimated treatment effect on each metric one by one, we can always easily find significant metrics, regardless of whether there is a real heterogeneity or not.

Many researchers attempted to address the "multiple comparison" problem, especially in the field of genomics. The most widely-adopted approach is the Benjamini– Hochberg (BH) method \cite{benjamini1995controlling}, which is easy to implement. However, BH method is known to control FDR under restrictive dependence assumptions. Its variation, Benjamini–Yekutieli (BY) method can work under arbitrary dependence assumptions \cite{benjamini2001control}, but it tends to be too conservative. 

Besides, due to the overly affluent amount of data for large online-controlled experimentation platform, scalability becomes a major concern. The goal of our work is to fill the gap by discover a computational-efficient statistical method that is powerful for detecting average treatment effect (ATE) under different dependence assumptions, while dealing with the potential "multiple comparison" problem by controlling the FDR below a pre-determined level.

To address the above problems, we propose \textbf{ATE-dBH} method, based on the dependence-adjusted Benjamini–Hochberg procedure (dBH) and additional approaches to improve computational efficiency. Using both simulation and real-experiment data, we demonstrate that dBH is more powerful than the original BH method, and have a better performance in large-scale A/B testing scenario compared to other FDR-control algorithms, such as knockoff \cite{candes2016panning}, adopted by Snap \cite{xie2018false}.

Here is a summary of our contributions:
\begin{itemize}[topsep=0pt]
    \item We establish ATE detection problem as  FDR control problem and integrate it with the potential outcome framework into the setting of large-scale online controlled experiments.
    \item We apply two FDR-control procedures (BH and dBH) to our hypothesis testing process for detecting significant ATE and share insightful comparisons of the two methods based on simulation and empirical data.
    \item We propose methods to improve the computation efficiency and discuss the scalability of dBH method in detail.
    \item We use our experimentation scenario to evaluate quantitatively and qualitatively different FDR control methods for multiple comparisons to control (MCC) problem. 
\end{itemize}


\section{Preliminaries}

\subsection{Average Treatment Effects}
Under the framework of Rubin causal model, online metrics can be rigorously defined. Let $Y_m^{i}(g_i)$ be the potential outcome for user $i$ of metric $m$, where $g_i = 0$ indicates the user $i$ is in control group, and $g_i = 1, 2, \ldots$ indicate assignment in different treatment groups. Without loss of generality, consider the case with only one treatment group and control group. To measure unit-level causal effect, $\tau_m^{i} = Y_m^{i}(1) - Y_m^{i}(0)$ is the causal effect for taking the treatment, and the average $\tau_m$ over all users is defined as "Average Treatment Effect" (ATE). However, $\tau_m$ is not observable since we cannot directly measure every $Y_m^{i}(1)$ and $Y_m^{i}(0)$ at the same time, and this dilemma is well known as the “fundamental problem of causal inference” \cite{holland1986statistics}.

Although we cannot observe $\tau_m$, with proper randomization guarantees \textit{Ceteris paribus} (other things equal), it is both intuitive and rigorously provable under Rubin’s causal model that the difference $\hat{\tau}_m$ between average of observed outcome over users in treatment group, denoted as $\bar{Y}_m^{obs}(1)$ and users in control group, $\bar{Y}_m^{obs}(0)$, would be an unbiased estimator of $\tau_m$,
$$
\tau_m = \mathbf{E}(\hat{\tau}_m) = \mathbf{E} (\bar{Y}_m^{obs}(1) - \bar{Y}_m^{obs}(0))
$$
given that the following two assumptions hold:
\begin{itemize}
    \item Stable unit treatment value assumption (SUTVA):
    Treatment applied to one user does not affect the outcome of other users (no interference).
    \item Unconfoundedness: 
    $g_i \perp (Y_m^{i}(1), Y_m^{i}(0)) \mid X_i$, where $X_i$ is a set of pre-treatment variables for $i^{th}$ user, such as gender, age, etc.
\end{itemize}



\subsection{Multiple comparison problem}
In industry, there are many built-in tools for practitioners to carry out large-scale A/B testing easily, but in fact many non-statisticians will adopt a naive approach that leads to spurious discovery of ATE. Suppose for an experiment, there are 1 treatment groups, 1 control group and 20 metrics. It seems intuitive to carry out a total number of 20 two-sample z test to compare the difference of each metric between treatment group to control group respectively, but it will suffer from the so-called "multiple comparison problem":

Assume the significance level is 0.05, then the probability of observing at least one significant result just due to random chance would be:
$$
\begin{aligned}
    P(\text{at least one significant result}) &= 1 - P(\text{no significant results}) \\
    &= 1 - (1 - 0.05)^{20} \approx 0.642
\end{aligned}
$$

Therefore, with 20 tests being considered, we have a $64.2\%$ chance of observing at least one significant result, even if there's no significant ATE to detect across all metrics.


To mitigate this problem, in our large-scale A/B testing settings, we consider controlling the term, "false discovery rate" (FDR), which is defined as:
$$
FDR  = E(Q)
$$
where $Q$ is the proportion of false discoveries among all the discoveries (rejection of null hypothesis).  

The same problem can be extended to the setting where there are multiple treatment groups and one single control group - a setting known as \textit{multiple comparisons to control} (MCC) problem, where dependence arises another major concern. 

In MCC, consider obtaining an average treatment effect estimator for every treatment group compared to the pre-determined control group. Then, all ATE estimators are correlated because they are linked together through the average of control group. In practice, dependence can be difficult to handle in many commonly used FDR control methods (see Section 5 for more details), not to mention the possibility of high correlation among metrics within one group itself, e.g., the number of daily active users and the number of weekly active users. 

To sum up, in multiple comparisons problem, two main issues demand attention: controlling FDR level and incorporating information from dependence.
 






\section{Detection of ATE with FDR control} 
\input{integration_testing_causality}
In this section, we first present how Benjamini–Hochberg procedure (BH) can be applied to detect ATE for online metrics in large-scale A/B testing platform and introduce the dependence-adjusted Benjamini–Hochberg procedure (dBH). We also include extensive simulation and empirical result to compare the two methods.

For simplicity of illustration, suppose for an online controlled experiment, we have treatment group $g = 1$, $g = 2$ and control group $g = 0$, and there are a number of $m$ metrics; then we can calculate the mean of each metric $i \in \{1,...,m\}$ for both treatment and control groups, denoted as $\bar{Y_i}^{obs}(2)$, $\bar{Y_i}^{obs}(1)$ and $\bar{Y_i}^{obs}(0)$. Then, we want to test a sequence of null hypotheses $H_1, ..., H_m$ to detect whether the mean for each metric from treatment group $g = 1$ and control group $g = 0$ are significantly different; and the same steps to test $H_{m+1}, ..., H_{2m}$ for  comparison between treatment group $g = 2$ and control group $g = 0$.

Now, for comparison of group means, we consider adopting z-test, which is reasonable in practice because in large-scale experiment the sample size is typically more than magnitude of $10^6$, and the means should approximate the normal distribution given Central Limit Theorem \cite{le1986central}. Noted that the central limit theorem only assumes finite variance, which almost always applies in online experimentation \cite{deng2014statistical}, and the speed of convergence to normal distribution can be quantified with an upper bound by using Berry-Esseen theorem \cite{berry1941accuracy,esseen1942liapunov}. 

Therefore, for hypotheses $H_1, ..., H_m$, we can calculate z-statistics for each metric $i \in \{1,...,m\}$ by $z_i(1) = \frac{\bar{Y_i}^{obs}(1) - \bar{Y_i}^{obs}(0)}{\sqrt{\sigma_i^2(1)/n_1 + \sigma_i^2(0)/n_0}}$, where $\sigma_i^2(1)$, $\sigma_i^2(0)$, $n_1$, $n_0$ are standard deviation and sample size for metric $i$ of treatment group $g = 1$ and control group $g = 0$ respectively. We denote the vector of z-statistics as $\mathbf{Z(1)} = [z_1(1), ..., z_m(1)]$; and the same for treatment group $g = 2$, with a z-statistics vector $\mathbf{Z(2)} = [z_1(2), ..., z_m(2)]$. Then we concatenate the two vectors together to get the vector of z-statistics for the entire experiment $\mathbf{Z} = [\mathbf{Z(1)},\mathbf{Z(2)}]$. $\mathbf{Z}$ can be further appended with $\mathbf{Z(3)},\mathbf{Z(4)}, \ldots$, if there are multiple treatment groups.


\subsection{BH and BY method}
In order to deal with the multiple testing problem and in the meantime reduce the conservativeness, Benjamini and Hochberg proposed Benjamini-Hochberg (BH) procedure \cite{benjamini1995controlling} to control FDR. BH procedure is known to control FDR if the test statistics are independent or obey the positive regression dependence on a subset property introduced in \cite{benjamini2001control}, but is not universally valid. However, due to its simplicity, practitioners still commonly default to this uncorrected BH method, choosing to forego theoretical guarantees and hope for the best \cite{fithian2020conditional}. 

BH procedure works as follows: Suppose that we have p-values from $m$ hypothesis testings $H_1, \ldots , H_m$, first rank p values on ascending order $p(1), ..., p(m)$, then find the largest $k$ such that $p(k) \leq \frac{k}{m}q$, and reject all null hypothesis $H_i$ for $i \leq k$. By doing so, it theoretically guarantees that FDR is controlled under $q$ in those restrictive dependence scenarios \cite{xie2018false}. 


Based on BH procedure, we propose following ATE-BH method:
\begin{itemize}[topsep=0pt]
    \item Step 1: Calculate the vector of z-statistics $\mathbf{Z}$ for this experiment based on observed data.
    \item Step 2: Transform p-values from every value of $z$ in $\mathbf{Z}$: 
    $p = 2 \Phi(-|z|)$ if two-sided test, $p = 1 - \Phi(z)$ if right-sided test, and $p = 1 - \Phi(z)$ if left-sided test. $\Phi(\cdot)$ is CDF for standard normal distribution.
    \item Step 3: Apply BH procedure on these p-values to finalize the list of selected significant metrics where null hypotheses are rejected.
\end{itemize}

To apply BY procedure, the only refinement is to modify the threshold and find the largest $k$ such that: $p(k) \leq \frac{k}{m \cdot c(m)}q$, where $c(m) = \sum_1^m 1/i$; and the rest steps remain the same. BY procedure can work under arbitrary dependence structure but is known to be very conservative \cite{korthauer2019practical}, leading to a high rate of false negative and low statistical power.

\subsection{Dependence BH Method}
"Conditional calibration for false discovery rate control under dependence" is a recently proposed FDR control method \cite{fithian2020conditional}. The technical idea is to decompose the FDR according to the additive contribution of each hypothesis, and use conditional inference to calibrate a separate rejection rule for each hypothesis adaptively, that can control its FDR contribution directly. Specifically, this idea can be applied to BH method by calibrating a separate BH p-value cutoff for each hypothesis, and it is  called dependence-adjusted Benjamini–Hochberg procedure (dBH) \cite{fithian2020conditional}. 

\subsubsection{ATE-dBH method}
In order to better detect significant metrics under dependence while adaptively control FDR, we propose the following procedure, \textbf{ATE-dBH} method:

\begin{itemize}[topsep=0pt]
\item Step 1: Calculate the vector of z-statistics $\mathbf{Z}$, and its corresponding correlation matrix $\Sigma$, based on observed data.

\item Step 2: Transform p values from every value of $z$ in $\mathbf{Z}$: 
$p = 2 \Phi(-|z|)$ if two-sided test, $p = 1 - \Phi(z)$ if right-sided test, and $p = 1 - \Phi(z)$ if left-sided test. $\Phi(\cdot)$ is CDF for standard normal distribution.
    
\item Step 3: Calculate the set of rejected hypotheses from BH method $\mathcal{R}^{\mathrm{BH}(\gamma \alpha)}$. Here $\alpha$ is the pre-set significance level, and $\gamma$ is a tuning parameter for FDR. We typically choose $\gamma = 1$ for one-sided test and $\gamma = 0.95$ for two-sided test.

\item Step 4: Calculate the set of rejected hypotheses from BH method $\mathcal{R}^{\mathrm{BH}(q_i)}$. Here $q_i = min(\{\alpha : i \in \mathcal{R}^{\mathrm{BH}(\alpha)} \})$, representing the level at which $H_i$ is barely rejected. $q_i$ is also known as storey's q-value \cite{storey2003positive}.

\item Step 5: For each $p_i$, calculate the conditional expectation, by integrating over $\{(z, S_i + \Sigma_{i,-i}z), z \in \mathbbm{R}\}$:
$$
g_i^*(q_i; S_i) = \mathbb{E}_{0}\left[\frac{\mathbf{1}\left\{p_{i} \leq q_i \mathcal{R}^{\mathrm{BH}(q_i)}/n \right\}}{\mathcal{R}^{\mathrm{BH}(\gamma \alpha)}} \mid S_{i}\right] \leq \frac{\alpha}{n} 
$$
where $S_i = \mathbf{Z}_{-i} - \Sigma_{-i,i} \mathbf{Z}_i$, $n$ is the number of elements in $\mathbf{Z}$. 


\item Step 6: Return the rejection set $\mathcal{R} = \{i: g_i^*(q_i; S_i) \leq \alpha/n \}$, and we can draw the conclusion that the corresponding metric in the corresponding group is significant.

\end{itemize}



\subsubsection{Calculation of correlation matrix}\label{cov}
We can see that step 1 of ATE-dBH method requires the computation of covariance matrix $\Sigma$ of the vector of z-statistics $\mathbf{Z}$, which can be computational expensive in practice \cite{xiong2021covariance}. However, we discover that in large-scale A/B testing setting, we can reduce the cost drastically from calculating the covariance between metrics across all groups to only within control group $g = 0$. 

First, given $\mathbf{Z} = [z_{1}(1), \ldots,z_{m}(1),\ldots, z_{1}(g), \ldots, z_{m}(g)]$, the covariance between z-statistics of metric $m$ from group $g$ and $m'$ from group $g'$ is calculated as:
\begin{equation}
\resizebox{.95\hsize}{!}{$
cov(z_{m}(g),z_{m'}(g')) = 
\frac{cov(\bar{Y}_{m}^{obs}(g) - \bar{Y}_{m}^{obs}(0) , \bar{Y}_{m'}^{obs}(g') - \bar{Y}_{m'}^{obs}(0)) } {\sqrt{(\sigma^2_{m}(g)/n_g + \sigma^2_{m}(0)/n_0)(\sigma^2_{m'}(g')/n_{g'} + \sigma^2_{m'}(0)/n_0)}} $}
\notag
\end{equation}

For large-scale A/B Testing, since traffic is diverted randomly to different treatment and control groups, we can make the following two valid assumptions:

\begin{itemize}[topsep=0pt]
    \item For any metric $m$, there are no correlation between different groups $g \neq g'$.
    $$cov(\bar{Y}_{m}^{obs}(g) , \bar{Y}_{m}^{obs}(g')) = 0$$
    \item For any pair of different metrics $m$ and $m'$,  the covariance is the same across any different group $g$ and $g'$:
    $$\begin{aligned}
    cov(\bar{Y}_{m}^{obs}(g) , \bar{Y}_{m'}^{obs}(g)) &= cov(\bar{Y}_{m}^{obs}(g') , \bar{Y}_{m'}^{obs}(g'))\\
    &= cov(\bar{Y}_{m}^{obs}(0) , \bar{Y}_{m'}^{obs}(0))
    \end{aligned}$$
\end{itemize}

Thus, for any element of $\Sigma_{i,j}$, let $k = \lfloor i/m \rfloor$, $k' = \lfloor j/m \rfloor$, $q = i \pmod m$ and $q' = j \pmod m$, the value of $\Sigma_{i,j}$ is:

\[
\Sigma_{i,j} =
C_{i,j}
\begin{cases}
\frac{1}{C_{i,j}} & \text{if $k= k' \textit{ and } q = q'$} \\
\sigma_q^2(0)/n_0 & \text{if $k\neq k' \textit{ and } q = q'$} \\
2 cov(\bar{Y}_{q}^{obs}(0) , \bar{Y}_{q'}^{obs}(0))  & \text{if $k=k' \textit{ and } q \neq q'$} \\
cov(\bar{Y}_{q}^{obs}(0) , \bar{Y}_{q'}^{obs}(0)) & \text{if $k \neq k' \textit{ and } q \neq q'$} 
\end{cases}
\]
where $C_{i,j} = 1/ \sqrt{(\sigma^2_{q}(k)/n_k + \sigma^2_{q}(0)/n_0)(\sigma^2_{q'}(k')/n_{k'} + \sigma^2_{q'}(0)/n_0)}$

Therefore, we can see that the calculation of $\Sigma$ can be reduced to only calculating the covariance between metrics within control group $0$. Detailed derivation is included in appendix.

\subsection{Simulations}
To demonstrate the statistical power and FDR control of ATE-dBH method, we compare it with classical BH and BY methods, and carry out simulations in different correlation structures. 

We assume a multivariate normal distribution with $m = 50$ and $\mathbf{Z} \sim \mathcal{N}_m (\mu, \Sigma)$, where 
$\mu_1, ..., \mu_5 = \mu^{*}, \mu_6, ..., \mu_{50} = 0$
and $\mu^*$ is signal size. For the covariance structures, we consider the following 3 types:
\begin{itemize}[topsep=0pt]
    \item Compound symmetry, where correlations are presumed to be the same for each metrics. Here we assume $\Sigma_{i,j} = 0.8^{1(\mid i - j \mid > 0)}$.
    \item Toeplitz covariance structure, where correlation values are decreasing for metrics that are increasingly far away. Here we assume $\Sigma_{i,j} = 0.8^{\mid i - j \mid}$.
    \item Block dependence, where the corresponding metrics are mostly dependent within but not between blocks. Here we assume $\Sigma_{i,j} = 0.8 \cdot \mathbf{1}(\lceil i/5 \rceil = \lceil j/5 \rceil)$.
\end{itemize}

\begin{figure*}[h!]
  \centering
  \includegraphics[width=1.8\columnwidth,height = 0.5\paperheight]{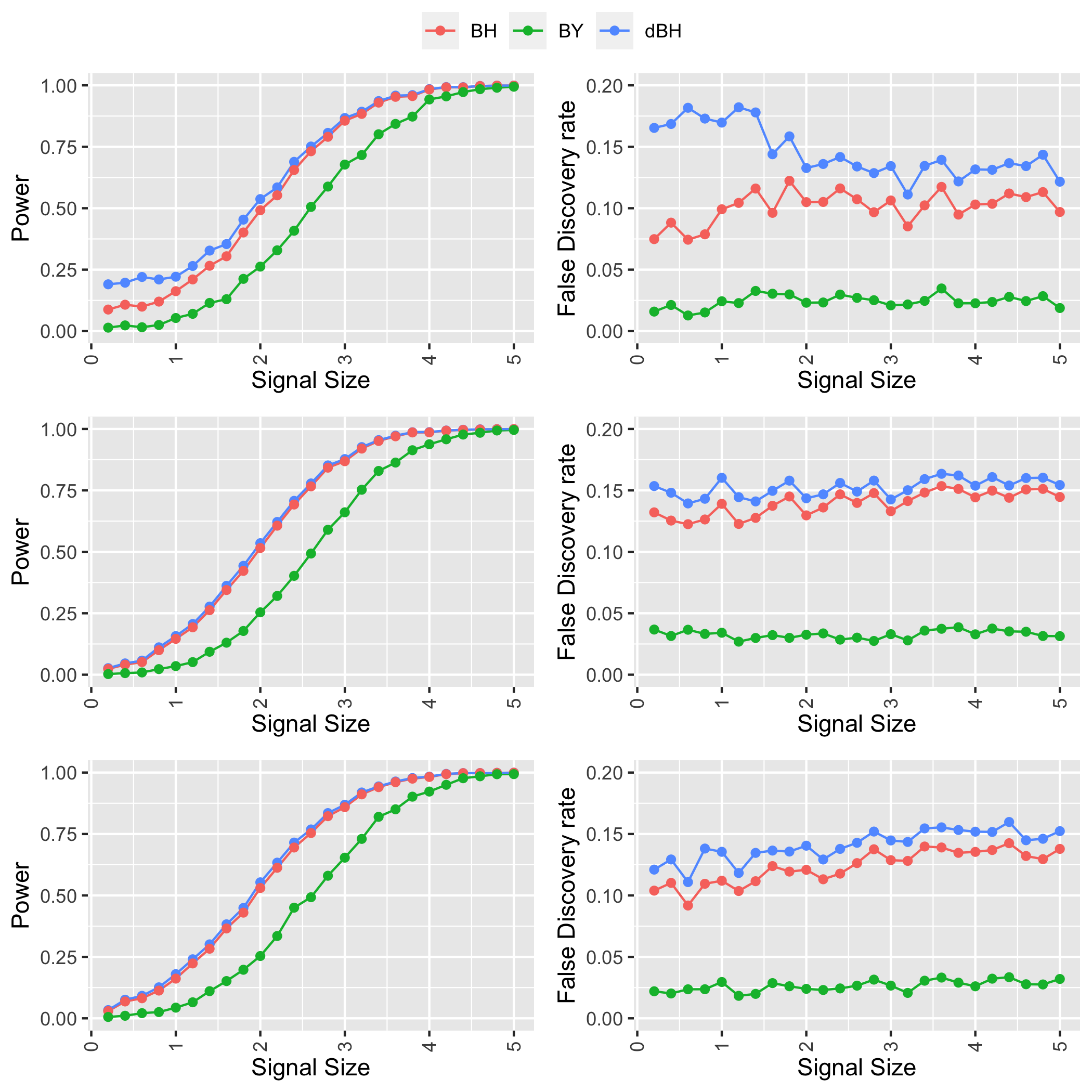}
  \caption{Estimated FDR and power for one-sided testing under $\alpha = 0.2$}
  \label{fig:sim1}
\end{figure*}

\begin{figure*}[h!]
  \centering
  \includegraphics[width=1.8\columnwidth,height = 0.5\paperheight]{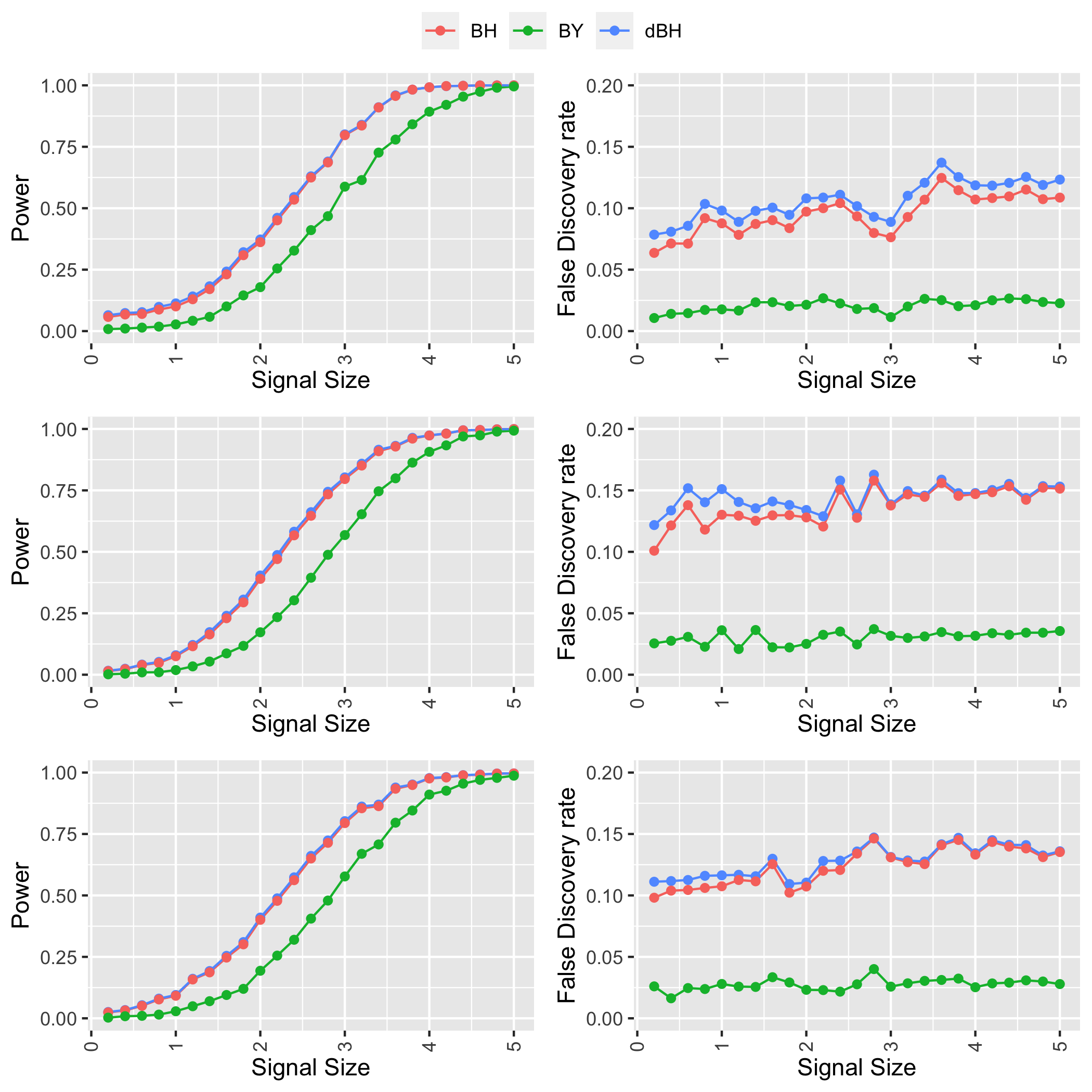}
  \caption{Estimated FDR and power for two-sided testing under $\alpha = 0.2$}
  \label{fig:sim2}
\end{figure*}

We perform both one-sided testing and two-sided using $dBH(\gamma\alpha)$, $BH(\alpha)$ and $BY(\alpha)$, where we choose $\gamma = 1$ for one-sided testing and $\gamma = 0.95$ for two-sided testing. We set the significance level $\alpha = 0.2$ and increase the signal size $\mu^*$ gradually to estimate their power and FDR respectively. At each step we run each of the above methods on 1000 individual samples. See Figure \ref{fig:sim1} and Figure \ref{fig:sim2}.

We observe that dBH improves the power of BH and BY at almost every given signal size $\mu^*$, while still maintain FDR under predetermined level $\alpha = 0.2$ for both one-sided and two-sided testing. From simulation, the average of power improvement for dBH over BH is approximately $3\%$.





\subsection{Empirical Result}
\begin{figure}[h!]
    \centering
    \includegraphics[width=\linewidth,height = 0.25\paperheight]{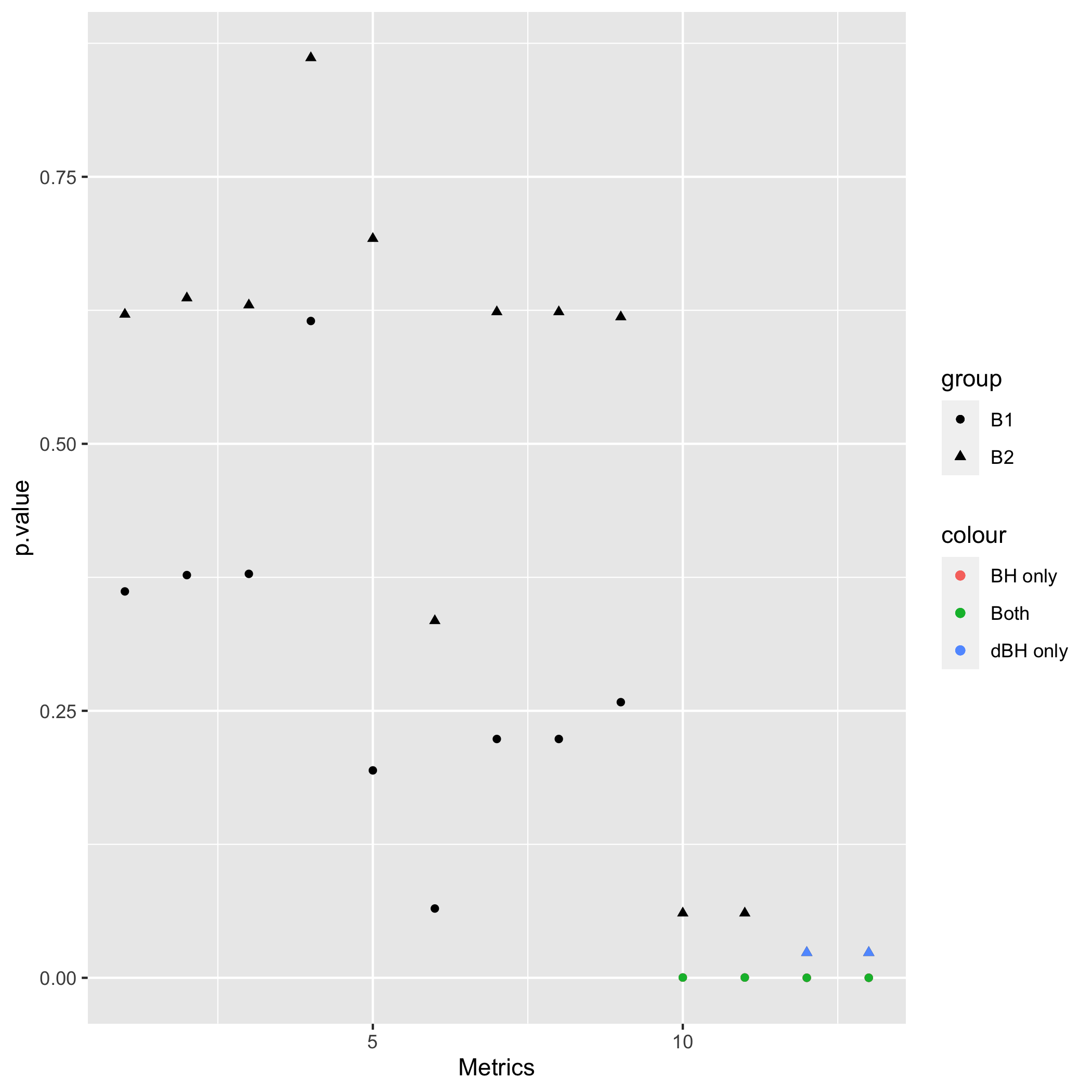}
    \caption{ Real A/B Testing results from using dBH and BH. Here we set the pre-specified FDR control level q = 0.1.}
    \label{fig:emp1}
\end{figure}

\begin{figure}[h!]
  \centering
  \includegraphics[width=\linewidth, height = 0.25\paperheight]{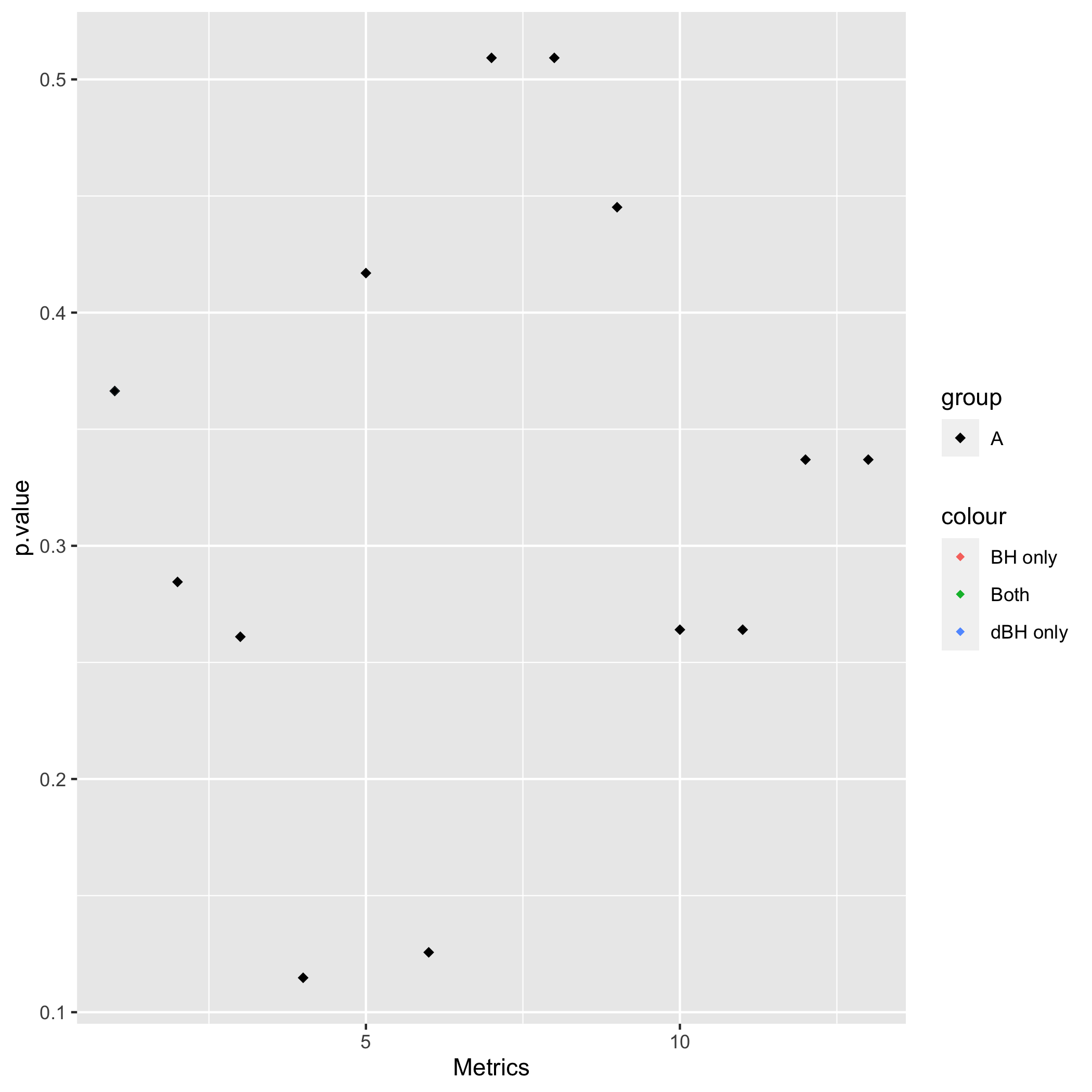}
  \caption{ Real A/A Testing results from using dBH and BH. Here we set the pre-specified FDR control level q = 0.2.}
  \label{fig:emp2}
\end{figure}

We apply our ATE-dBH method as well as BH method on two real experiment datasets. Results are displayed in Figure \ref{fig:emp1} and \ref{fig:emp2}.

In the first experiment, we have different treatment groups $B_1$ and $B_2$, and we can see from Figure \ref{fig:emp1} that dBH detects more significant metrics than BH and BY. The extra metrics detected by dBH are metric 12 and 13 for group $B_2$, which represent the number of clicks and click-through-rate for exposed users. We retrospectively understood that these metrics should be significant, since both treatment are designed to encourage clicks by assigning users different rewards for clicks.

In the second experiment, we conduct a A/A test, where there are control groups that are not given any treatments. In this case, Figure \ref{fig:emp2} demonstrates that both dBH and BH detect no significant metrics, and this should hold for A/A test because both control groups should be homogeneous.

\input{discussion.tex}



\section{Conclusion and Future Work}
We first propose ATE-BH method for detecting treatment effects in large-scale A/B testing setting based on traditional BH method, and further propose ATE-dBH method that can separately calibrate a data-dependent p-value rejection threshold for each hypothesis. From simulation data under both one-sided and two-sided testing, we demonstrate that dBH is more powerful than BH under various dependence scenarios, while still controlling false discovery rate under a pre-determined level. Real-world examples are also presented where ATE-dBH is able to detect more significant metrics.

In the discussion section, we offer extensive comparisons of dBH to other recently proposed FDR control methods, such as knockoff, and AdaPT. In the large scale experimentation scenario, dBH outcompetes these methods given that it is computationally efficient and can work better when the significant metrics are sparse.

For further research, we will consider the extension from average treatment detection to heterogeneous treatment detection. For example, users from different cities might react differently to the same treatment. Thus, more than detecting ATE, capturing such heterogeneity of treatment effects will also be useful. We will consider transformed outcome regression \cite{xie2018false} or conditional mean regression \cite{arminger1999mixtures} for estimating heterogeneous treatment effect and integrating them with dBH method in future work.

\section{Acknowledgement}
We thank Dr. Lihua Lei from Stanford University for insightful discussions on the pros and cons for dependence-adjusted BH procedure. We thank the engineering team for providing valuable data sets of online controlled experiments.

\appendix

\section{calculation of covariance matrix}

Assume we have a number of $g$ treatment groups, one control group and a number of $m$ metrics. Then we carry out z-tests to compare every treatment group to the control group for every metrics, which leads to a number of $gm$ hypothesis testing. 

Thus, we define $z_{k}(j)$ to be the z-score derived from z-test of metric $k$ between treatment group $j$ and control group $0$:
$$
z_{k}(j) = \frac{\bar{Y}_k^{obs}(j) - \bar{Y}_k^{obs}(0)}{\sqrt{\sigma^2_{k}(j)/n_k + \sigma^2_{k}(0)/n_0}}
$$
where $\bar{Y}_k^{obs}(j)$, $\sigma^2_{k}(j)$ and $n_k$ are observed sample mean, variance and sample size for treatment group $j$ of metric $k$; $\bar{Y}_k^{obs}(0)$, $\sigma^2_{k}(0)$ and $n_0$ are corresponding value for control group.

Then we can define the vector of z-scores, denoted as $\mathbf{Z} = [z_{1}(1), \ldots,z_{m}(1),\ldots, z_{1}(g), \ldots, z_{m}(g)]$.

To calculate the covariance matrix $\Sigma$ of $\mathbf{Z}$, we need to compute the covariance between every pair of $z_{m}(g)$ and $z_{m'}(g')$.

\begin{equation}
\resizebox{.95\hsize}{!}{$
cov(z_{m}(g),z_{m'}(g')) = 
\frac{cov(\bar{Y}_{m}^{obs}(g) - \bar{Y}_{m}^{obs}(0) , \bar{Y}_{m'}^{obs}(g') - \bar{Y}_{m'}^{obs}(0)) } {\sqrt{(\sigma^2_{m}(g)/n_g + \sigma^2_{m}(0)/n_0)(\sigma^2_{m'}(g')/n_g' + \sigma^2_{m'}(0)/n_0)}} $}
\notag
\end{equation}

Since for A/B Testing, traffic are diverted randomly to different treatment and control group, we assume the following:

1) For any metric $m$, there are no correlation between different groups $g \neq g'$.
$$
cov(\bar{Y}_{m}^{obs}(g) , \bar{Y}_{m}^{obs}(g')) = 0 
$$

2) For any pair of different metrics $m$ and $m'$,  the covariance is the same across any different group $g$ and $g'$ .
$$
\begin{aligned}
cov(\bar{Y}_{m}^{obs}(g) , \bar{Y}_{m'}^{obs}(g)) &= cov(\bar{Y}_{m}^{obs}(g') , \bar{Y}_{m'}^{obs}(g'))\\
&= cov(\bar{Y}_{m}^{obs}(0) , \bar{Y}_{m'}^{obs}(0))
\end{aligned}
$$

There are four cases:

1) if $g = g'$ and $m = m'$,
\begin{equation*}
\resizebox{1.05\hsize}{!}{$
\begin{split}
cov(z_{m}(g),z_{m'}(g')) 
&= \frac{cov(\bar{Y}_{m}^{obs}(0) , \bar{Y}_{m'}^{obs}(0)) + cov(\bar{Y}_{m}^{obs}(g) , \bar{Y}_{m}^{obs}(g))} {\sqrt{(\sigma^2_{m}(g)/n_g + \sigma^2_{m}(0)/n_0)(\sigma^2_{m'}(g')/n_g' + \sigma^2_{m'}(0)/n_0)}}\\
&= \frac{\sigma^2_m(0)/n_0 + \sigma^2_m(g)/n_g} {\sqrt{(\sigma^2_{m}(g)/n_g + \sigma^2_{m}(0)/n_0)(\sigma^2_{m'}(g')/n_g' + \sigma^2_{m'}(0)/n_0)}} = 1 
\end{split}$}
\notag
\end{equation*}

2) if $g \neq g'$ and $m = m'$,
\begin{equation*}
\resizebox{1.0\hsize}{!}{$
\begin{split}
cov(z_{m}(g),z_{m'}(g'))  
&= \frac{cov(\bar{Y}_{m}^{obs}(0) , \bar{Y}_{m'}^{obs}(0)) } {\sqrt{(\sigma^2_{m}(g)/n_g + \sigma^2_{m}(0)/n_0)(\sigma^2_{m'}(g')/n_g' + \sigma^2_{m'}(0)/n_0)}}\\
&= \frac{\sigma^2_m(0)/n_0} {\sqrt{(\sigma^2_{m}(g)/n_g + \sigma^2_{m}(0)/n_0)(\sigma^2_{m'}(g')/n_g' + \sigma^2_{m'}(0)/n_0)}}
\end{split}$}
\notag
\end{equation*}

3) if $g = g'$ and $m \neq m'$,
\begin{equation*}
\resizebox{1.0\hsize}{!}{$
\begin{split}
cov(z_{m}(g),z_{m'}(g')) 
&= \frac{cov(\bar{Y}_{m}^{obs}(0) , \bar{Y}_{m'}^{obs}(0)) + cov(\bar{Y}_{m}^{obs}(g) , \bar{Y}_{m'}^{obs}(g)) } {\sqrt{(\sigma^2_{m}(g)/n_g + \sigma^2_{m}(0)/n_0)(\sigma^2_{m'}(g')/n_g' + \sigma^2_{m'}(0)/n_0)}}\\
&= \frac{2 cov(\bar{Y}_{m}^{obs}(0) , \bar{Y}_{m'}^{obs}(0))} {\sqrt{(\sigma^2_{m}(g)/n_g + \sigma^2_{m}(0)/n_0)(\sigma^2_{m'}(g')/n_g' + \sigma^2_{m'}(0)/n_0)}}
\end{split}$}
\notag
\end{equation*}

Given assumption 2 that \text{\footnotesize$ cov(\bar{Y}_{m}^{obs}(0) , \bar{Y}_{m'}^{obs}(0)) = cov(\bar{Y}_{m}^{obs}(g) , \bar{Y}_{m'}^{obs}(g))$}.

4) if $g \neq g'$ and $m \neq m'$,
\begin{equation*}
\resizebox{1.05\hsize}{!}{$
\begin{split}
cov(z_{m}(g),z_{m'}(g')) 
&= \frac{cov(\bar{Y}_{m}^{obs}(0) , \bar{Y}_{m'}^{obs}(0)) } {\sqrt{(\sigma^2_{m}(g)/n_g + \sigma^2_{m}(0)/n_0)(\sigma^2_{m'}(g')/n_g' + \sigma^2_{m'}(0)/n_0)}}
\end{split}$}
\notag
\end{equation*}

\bibliographystyle{ACM-Reference-Format}
\bibliography{sample-base}

\end{document}

%% file: integration_testing_causality.tex
\subsection{Integration of FDR Control and Causal Inference}

%% file: discussion.tex
\section{Scalability and Computational Complexity}
The computational cost resides on two parts: the computational complexity of dBH algorithm itself and the computational cost of computing the inputs to dBH procedure - the estimated covariance matrix and the multivariate normal estimator. 

Given the inputs - the observed statistic $\mathbf{z}\in \mathbb{R}^{m}$ and the estimated covariance matrix $\Sigma \in \mathbb{R}^{m\times m}$, time complexity for dBH procedure is $\mathcal{O}(m^2$log$m)$, where $m$ is the dimension of our multivariate test statistic \cite{fithian2020conditional}; in our empirical example, $m = 26$. In cases above, on Apple 15" MacBook Pro 2.8GHz Intel Core i7, the task can be finished within seconds using the 
c++ implementation. For the computation of $\mathbf{z}$ statistics, suppose there are $N$ observed units at total. The calculation for $\mathbf{z}$ is thus $\mathcal{O}(N)$, which is still computationally feasible for a large-scale experimentation platform.  

The calculation of covariance matrix $\Sigma$ of $\mathbf{z}$ counts for most of the computational cost. Theoretically, dBH would be valid as long as the Frobenius norm of the difference between the estimated covariance matrix and the true covariance matrix asymptotically converges to zero \cite{li2021}. Therefore, to further reduce the computational costs in estimating the covariance matrix following the steps in Section \ref{cov} and Appendix A, covariance estimation can be calculated across grouped observed units \cite{xiong2021covariance}, instead of individual observed data points. Similar technique can be applied elsewhere, and practitioners can employ the covariance estimation methods that suit well their need. 

\section{Discussion}
In the setting of our current study, the vector of z-statistic $\textbf{Z}$ used for testing $H_i: \mu_i =0$ is known, and an estimated covariance matrix can be obtained. In this section, we introduce several other procedures that are studied and widely used in the literature of multiple testing under dependence, and offer comparisons to our ATE-dBH method.
\subsection{Knockoff Procedure} 
Knockoff procedure can be used for controlling FDR \cite{barber2015controlling}\cite{candes2016panning}\cite{xie2018false}. Traditionally, knockoff procedure aims to control the false discovery rate in the linear regression setting: $$\mathbf{y} = \mathbf{X}\mathbf{\beta} + \mathbf{\epsilon}$$
$\mathbf{y}\in \mathbb{R}^n$ is the observed response vector, $\mathbf{X}\in \mathbb{R}^{n\times p}$ is the known design matrix, $\mathbf{\beta}\in \mathbb{R}^{p}$ is the unknown coefficient vector and $\mathbf{\epsilon} \in \mathbb{R}^{n} \sim \mathcal{N}(0,\sigma^2\mathbf{I})$ is the unobserved noise vector. The knockoff procedure constructs "knockoff" variables using the original design matrix $\mathbf{X}$ to incorporate the correlation structure of the original features, calculate a statistic with both sufficiency and anti-symmetry property, and finally make discovery based on data-dependent threshold for the statistics. The "knockoff" procedure selects the set of variables whose regression coefficients are away from zero with higher statistical power, and a theoretically guaranteed statistics similar to FDR. Under the same significance level $\alpha$, knockoff usually enjoys higher statistical power than methods such as BH and BY, as correlation information is employed. 

While "Knockoff" procedure has been extensively studied for variable selection in the setting of linear regression, it can also be adapted to the setting of hypothesis testing using the multivariate normal estimators. In the current study, we considered two procedures that transform the hypothesis testing in MCC problem into the settings where "Knockoff" procedure can be directly used.

Li and Fithian \cite{li2021} have delineated the correspondence between "Knockoff" method and multivariate normal estimator formulations, simply by transforming the vector of z-statistics and the estimated covariance matrix to obtain a pseudo-design matrix and pseudo-response vector. A "whitened" estimator is obtained for inference using the proposed procedure. Then the multiple testing procedure for this whitened multivariate normal estimator can be conducted using any state-of-art implementation of knockoff methods. This method is known as \textit{"Knockoff-Whiteout"}. 

However, in this proposed procedure, there are additional noise terms $\epsilon$ that are devised for de-correlation, but they disturb the original signals. The "whiteout" problem comes when the covariance matrix structure is "too far from diagonal" and large noise signals are added to obtain a covariance matrix that are suitable for the knockoff problem. We refer the readers to look at \cite{li2021} for more technical details.

In online experimentation platforms, most online A/B testing has less than $10\%$ metrics detected as significant (In fact a $1\%$ change is quite rare for many key metrics), and the covariance matrices usually contain very large eigenvalues - in our empirical A/B testing example, the largest eigenvalue is 8.7, suggesting the noise needed for "Knockoff-Whiteout" procedure will disturb the signals. In this case, the "whiteout" phenomenon makes knockoff almost unable to make any discovery. 

Besides "whiteout", another work from Barber and Candes \cite{barber2015controlling} proved that "knockoff" is a special case of the multiple testing procedure "SeqStep". This method is called \textit{"Knockoff-SeqStep"}. 

To illustrate the performance of knockoff methods compared to our "ATE-dBH" procedure, both simulation and application on empirical A/B testing dataset are conducted.

\subsubsection{Simulation}
Here we adopt the same setting as simulation in section 3.4, and perform both one-sided and two-sided testing using BH, dBH and Knockoff-SeqStep. Knockoff-Whiteout procedure is omitted as it adds much noise and makes almost zero rejections in our setting with small proportion of true signals. We carry out simulations under both significance levels $\alpha=0.2$ and $\alpha=0.05$. Signal sizes  are tuned so that BH has approximately 0.3 power in a separate Monte Carlo simulation. Each of the above methods is run on 1000 independent samples of z-statistics and p-values. 

\begin{figure}[h!]
  \centering
  \begin{subfigure}{8cm}
  \includegraphics[width=\textwidth]{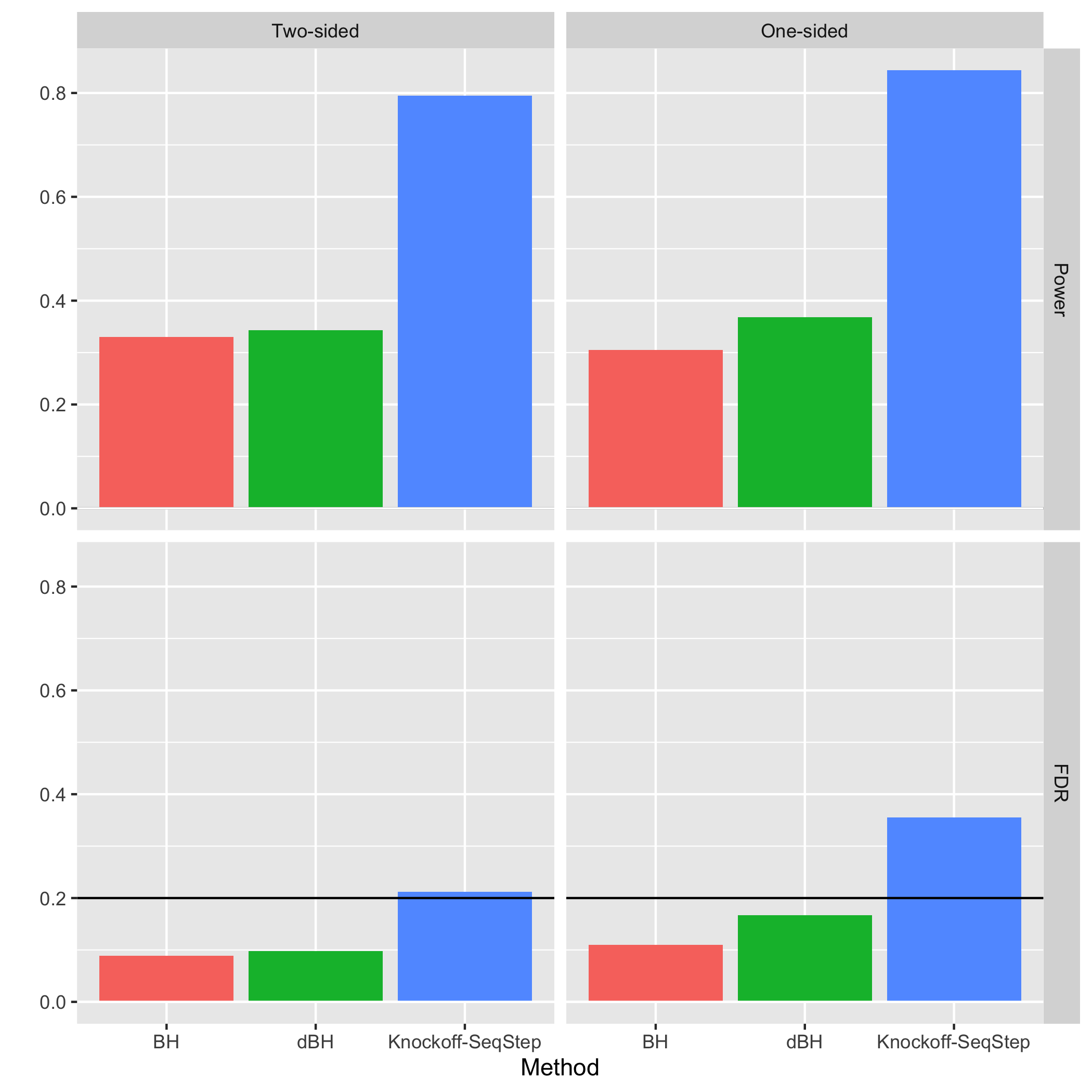}
  \subcaption{Compound symmetry correlation structure with $\alpha=0.2$}
  \label{fig:a02}
  \end{subfigure}
  
  
  \begin{subfigure}{8cm}
  \includegraphics[width=\textwidth]{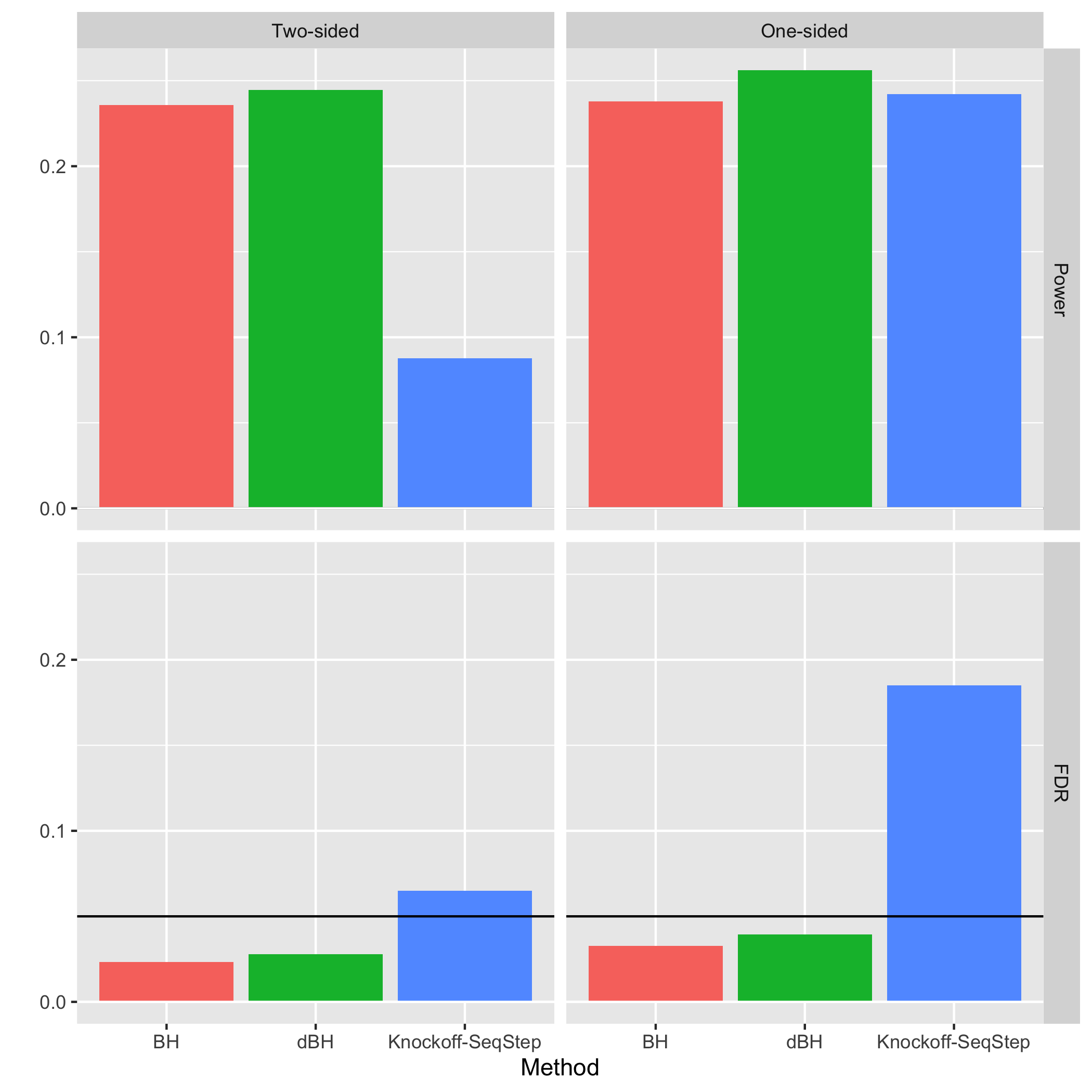}
  \subcaption{Compound symmetry correlation structure with $\alpha=0.05$}
  \label{fig:a005}
  \end{subfigure}
  \caption{Estimated FDR and power for Knockoff-SeqStep, dBH and BH under different significance levels.}
\end{figure}

FDR and power are estimated, presented in Figure \ref{fig:a02} and \ref{fig:a005}. 

Our simulation results reveal that when dependence presents, despite the larger power, "Knockoff-SeqStep" fails to control FDR under pre-specified level $\alpha$, especially when $\alpha$ is small. Besides, both dBH and BH method are able to control FDR under $\alpha$, but dBH demonstrates larger power in all cases.

\subsubsection{Empirical Result}
We also examined the performance of dBH, BH, and Knockoff methods using the empirical A/B testing datasets from section 3.5. The results reveal the failure of Knockoff-based methods on multiple testing where there is a small number of actual signals, and the dependencies among hypotheses are not negligible in FDR control. See Table \ref{table:fourmethods}. 
\begin{table}[h]
\begin{threeparttable}
\resizebox{0.48\textwidth}{!}{%
\begin{tabular}{lll}
    \rowcolor{gray}
     & Discovery ($\alpha$=0.05) & New Discovery ($\alpha$=0.1) \\ \hline
     dBH  & metric 10,11,12,13 & metric 25, 26 \tnote{i}\\
     BH  & metric 10,11,12,13 & None \\
     Knockoff-SeqStep  & None & None \\
     Knockoff-Whiteout  & None & None \\
     \hline
\end{tabular}
}
\begin{tablenotes}
      \small
      \item[i] metrics indexing of 25 and 26 are in treatment group $B_2$, which corresponds 
      
      to the same metrics indexing of 12 and 13 in treatment group $B_1$.
\end{tablenotes}
\end{threeparttable}
\vspace{0.2cm}
\caption{Discovery of significant metrics using dBH, BH and knockoff methods. "New Discovery" represents additional discoveries made at \textbf{$\alpha=0.1$} but not at \textbf{$\alpha=0.05$}.}
\label{table:fourmethods}
\end{table}


\subsection{Procedures Using Side Information}
There are FDR control procedures that can leverage contextual information from the covariates in exploring the dependence structure and predicting the proportion of true null hypotheses to improve the power of testing, e.g., AdaPT framework \cite{lei2016adapt}, structure-adaptive BH algorithm \cite{li2019multiple} and Independence Hypothesis Weight \cite{ignatiadis2017covariate}. 

AdaPT framework proceeds iteratively by proposing a rejection threshold, estimating the false discovery proportion for the proposed threshold, and either rejecting every hypothesis with $p-$value smaller than (or equal to) the threshold or continuing updating the threshold. The power of AdaPT depends on specific updating rules, even though FDR is guaranteed to be controlled in any case. To estimate the false discovery proportion, the authors proposed an EM algorithm that takes the context information: $x_i$ for each hypothesis, and $p-$value at each iteration $t$: $\tilde{p}_{t,i}$ to predict the conditional density of null p-values. In EM Algorithm, the featurization of each test poses challenge in  experimentation platform. On one hand, for every unit within each group, the dimension of observed covariates is large (there are usually thousands of covariates for large-scale A/B testings). Without any dimensionality reduction or other featurization modeling techniques, the feature space would be gigantic; moreover, these preprocessing techniques can be computationally costly themselves. Therefore, contextual information featurization and modeling is less favorable in our setting where each observed unit contains high dimensional covariates.

\subsection{Comments} 
As we can see from the simulation result in both section 3 and 4,  dBH and BH procedures consistently enjoy better power in experiments where the total number of rejections is relatively small at the pre-determined FDR level. The pattern also has a theoretical explanation, as \cite{fithian2020conditional} pointed that the knockoff method requires $\frac{1+A_t}{R_t} \leq \alpha$, where $A_t$ is the count of $W_j$-values smaller than $-t$ and $R_t$ is the count above t; noted that $W_j$ is the statistic calculated by the procedure and $t$ represents $t^{th}$ test. Thus, knockoff method cannot make any rejections unless it makes at least a number of $\frac{1}{\alpha}$ (e.g., 20 if $\alpha = 0.05$), and the result can be unstable if the number of rejections is on the order of $\frac{1}{\alpha}$. 

In short, dBH method can control FDR in MCC problem while enjoy power gain by using the information from the dependence structure, and calibrating for dependencies among hypotheses. Our simulation and experimentation on empirical data set also validates the scalability and feasibility of dBH method in large-scale online controlled experiments. 
